**DROUGHT | CLIMATE CHANGE**

Why global climate change amplifies hydrological extremes

# Droughts in Germany

Axel Kleidon

*The warmer temperatures of global climate change strengthen the water cycle, evaporation and precipitation increase. But the extremes of heavy rain, floods, dry periods and droughts will also increase. How does this fit together? Simple physical considerations show which factors mainly regulate the strength of the water cycle in the Earth system, and how this determines water availability on land. This can be used to interpret the observed changes in the water balance in Germany and explain the increasing dryness in Germany.*

The last few years have seen exceptionally warm and dry summers in Germany, which are generally attributed to global climate change. But there have also been heavy rains, storms and floods, such as the disaster in the Ahr valley. How can it be that global climate change is intensifying both the droughts and the periods of excess water? What looks contradictory at first glance can be explained quite simply in physical terms. To do this, we will first look at how the water cycle is integrated into the workings of the Earth system before applying this to global climate change and interpreting the changes in the water balance of Germany.

In order to understand droughts, heavy rainfall and other aspects of hydrological extremes, we link these processes to the main actors of the hydrological cycle, precipitation and evaporation. Precipitation is easy to perceive and observe, even if it is highly variable in space and time. Evaporation, on the other hand, occurs gradually and continuously without us being aware of it. It removes water from the soil and returns it to the atmosphere: the soil dries out. Vegetation plays an important role here, as its root systems can extract water from deeper soil layers for evaporation - something that bare soil or asphalted surfaces cannot do. A dry period is a phase without precipitation in which evaporation can only take place if soil water is available and accessible. The hydrological cycle becomes more extreme when these two fluxes become more out of balance - in other words, when precipitation increases, periods of precipitation become shorter and dry periods become more intense and longer. But why is the hydrological cycle becoming more variable and more extreme with climate change?

To do so, we need to look at the main drivers and how they are integrated into the workings of the Earth system as a whole. The starting point for this is the phase transitions of water, in particular evaporation, i.e. the phase transition from liquid to gas, and the reverse direction, the condensation of water vapour into liquid water (Figure 1). The reference state is the thermodynamic equilibrium of saturated air. In this state, evaporation and condensation compensate each other locally, the phase transitions occur at the same temperature and the air is saturated with water vapour. There are no net fluxes of evaporation or precipitation, i.e. no changes in the water vapour content, no temporal variability and therefore no hydrological cycle.

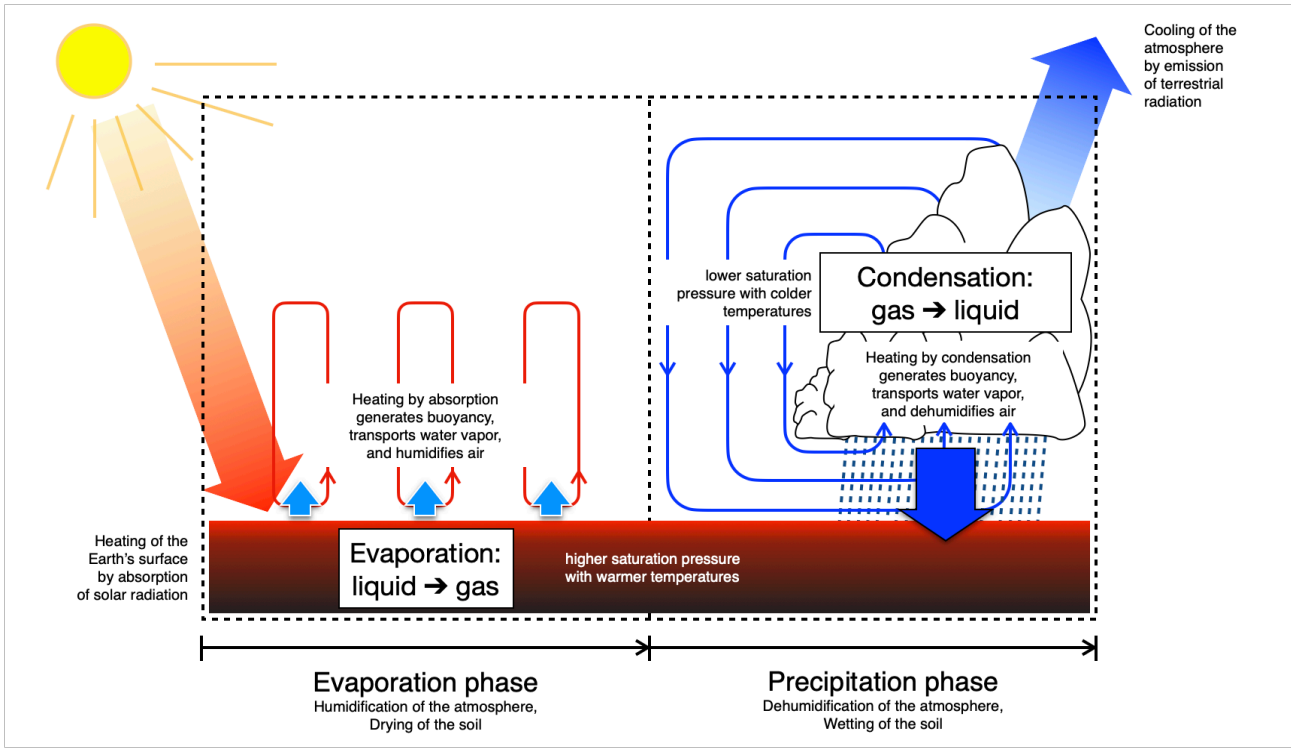

**FIG. 1 THE HYDROLOGICAL CYCLE**
*The hydrological cycle as a system far from thermodynamic equilibrium, in which heating and cooling by radiation and the phase transitions of evaporation and condensation are separated in space and time.*

The hydrological cycle therefore represents thermodynamic disequilibrium. This means that air is predominantly not saturated, i.e. its relative humidity is below 100 %, and the phase transitions are spatiotemporally separated. The spatiotemporal separation then looks like this: evaporation takes place at the heated Earth's surface, it brings water vapour into the atmosphere, humidifies it and reduces its disequilibrium. When precipitation takes place within the cooler atmosphere, condensed water vapour rains out, removing water, thus dehumidifying the air and creating the disequilibrium. This picture can be translated into a simple physical model. We use the model to describe the strength of the hydrological cycle and to estimate the effects of global climate change on the hydrological cycle (see box "Thermodynamics of the hydrological cycle" at the end of this document).

So what limits the intensity of the hydrological cycle? The key lies in the energy required to evaporate water. Evaporation requires a lot of energy: every kilogramme or litre of water needs around 2.5 MJ of energy to evaporate. This energy comes mainly from the absorption of solar radiation at the Earth's surface. Evaporation is therefore a key component of the surface energy balance. On a global average, it contributes more than half of the absorbed solar radiation to the atmosphere in the form of latent heat. Phase transitions of water and energy fluxes are thus closely linked in the Earth system.

## Air humidification through evaporation

The maximum possible evaporation rate is determined by a combination of energy availability, vertical transport and thermodynamics. Here we consider the case in which water availability of the soil is not a limiting factor. This leads us to the concept of equilibrium evaporation, also known as potential evaporation. It describes the upper limit of the maximum rate of water that can evaporate into the atmosphere. This can describe evaporation in humid regions, such as the wet tropics, very well, while in arid regions the water availability in the soil limits the actual evaporation rate, which is ultimately determined by the low precipitation. The comparison of potential evaporation and precipitation thus gives us a measure of the aridity of a region: in arid regions, potential evaporation is greater than precipitation, while in humid regions it is the other way round.

Potential evaporation is mainly determined by physics: Absorption of solar radiation heats the surface and thus the overlying air, creating buoyancy. This buoyancy transports the heated air into the atmosphere. This transports heat, resulting in a so-called sensible heat flow. It also transports the humidified air from the surface. This maintains evaporation and represents the latent heat flux. Maximum evaporation is achieved when the air remains saturated at the same time as it is heated at the surface, i.e. it is maintained in a state of thermodynamic equilibrium. This determines the partitioning into sensible and latent heat flux, the so-called equilibrium partitioning. This partitioning is set by the heat capacity of the air and the increase in saturation vapour pressure with temperature (see equation 2 in "Thermodynamics of the hydrological cycle"). While the heat capacity remains constant, the saturation vapour pressure increases exponentially with temperature. Thus, at warm temperatures, more of the heat input at the surface goes into evaporation instead of heating the air [1].

Buoyancy is also determined by thermodynamics. It is associated with work that is generated from the temperature difference between the Earth's surface and the atmosphere - as in a power plant that generates energy from the temperature difference between combustion and waste heat [2]. The more buoyancy is generated, the more heat is transported into the atmosphere, and the sensible and latent heat fluxes increase. The absorbed radiant energy from the sun and the radiation emitted downward by the atmosphere (the greenhouse effect) is then returned more and more into the atmosphere via the motion of air and less via the emission of thermal radiation of the surface. The surface therefore becomes cooler. This in turn reduces the efficiency with which power can be obtained from the vertical temperature difference. Therefore, there is a maximum of power that determines optimal temperatures and heat fluxes that agree very well with observations - atmospheric motion thus works at its maximum power limit [3].

Thermodynamics therefore acts twice to limit the evaporation rate and thus the input of water vapour into the hydrological cycle. It limits the total input of heat because it limits the power for vertical motion, and it determines the partitioning of this heat into heating and humidification of the air.

## Dehumidification through precipitation

The water vapour content in the atmosphere thus first increases. Water vapour is only removed after condensation has formed cloud droplets in the atmosphere and these rain out. Condensation mainly occurs when air rises, either through buoyancy, through large-scale motion associated with low-pressure systems, or when motion is forced to rise, such as on mountains. Air cools as it rises - this follows directly from the conservation of energy. As the potential energy increases during ascent, the thermal energy must decrease - so the rising air cools down. This leads to the adiabatic lapse rate, which describes the cooling of the air with height in the lower atmosphere very well. The water vapour therefore gets closer to saturation as the air rises.

When saturation and condensation occur, a self-reinforcing process starts: the latent heat of the water vapour is released, the air heats up, generating buoyancy and vertical motion. This draws in moist air from below, which is then also saturated and releases more heat. A cloud is formed. The released heat drives a heat engine and generates power - the power plant is the cloud. This power moves the air, causes moist convection and dehumidifies it. In detail, this happens as water droplets collide within the rising air, grow and become heavier until the buoyancy can no longer keep them in suspension. They fall down, create precipitation, thus dehumidifying the atmosphere and creating the disequilibrium.

If we now consider the hydrological cycle as a whole and on a climatological time scale, then it is a zero-sum game in terms of mass and energy. Evaporation and precipitation balance each other out. The energy used for evaporation is released again during condensation - this balances out as well. Thermodynamically, the hydrological cycle follows the second law, it produces entropy and reflects disequilibrium. This can be seen in the spatial separation of the processes: Energy is absorbed at the warmer surface of the Earth during evaporation, i.e. at lower entropy than when it is released through condensation in the cooler atmosphere. This produces entropy, so it is not a zero-sum game. Overall, this determines the intensity of the hydrological cycle because the boundary conditions are set by the energy balances - where the Earth heats up through the absorption of solar radiation and evaporation takes place, and where it is cooled through emission into space.

The fluxes, however, balance each other only on average. While evaporation occurs relatively continuously, precipitation events only take place when enough water vapour has accumulated in the atmosphere and condensation occurs. Precipitation is therefore more sporadic, more variable and involves higher rates than evaporation. In addition, there is the horizontal transport of water vapour within the atmosphere. This can bring evaporation and precipitation further out of balance in terms of space and time, transport the water vapour from the ocean to the land and separate regions into humid and arid areas in terms of their water availability.

## Water availability in Germany

What are the components of the hydrological cycle for Germany? Next, we look at these components for the climatological reference period of 1961-1990 (Figure 2). To do so, we use the HYRAS dataset of the German Weather Service [4], which spatially interpolates measurements from weather stations and covers more than the last 60 years.

While precipitation and solar radiation are directly part of the data set - after all, they are easily observable variables - evaporation from the surface is less well determined. For this we use the empirical estimate by Hargreaves [5], which determines the potential evaporation from the daily mean temperature and the amplitude of the diurnal cycle (see supplementary material). This temperature information is also available in the HYRAS dataset, while the energy balance-based approach from thermodynamics (see "Thermodynamics of the hydrological cycle", equation 2) also requires the downwelling longwave radiation emitted by the atmosphere, which is not routinely measured. Potential evaporation does not determine the actual water loss from the surface due to evaporation - for that we would need more detailed information, especially with regard to soil water availability and the vegetation cover. However, it does give us an idea of how much water could be lost through evaporation and therefore is a measure of water availability and dryness.

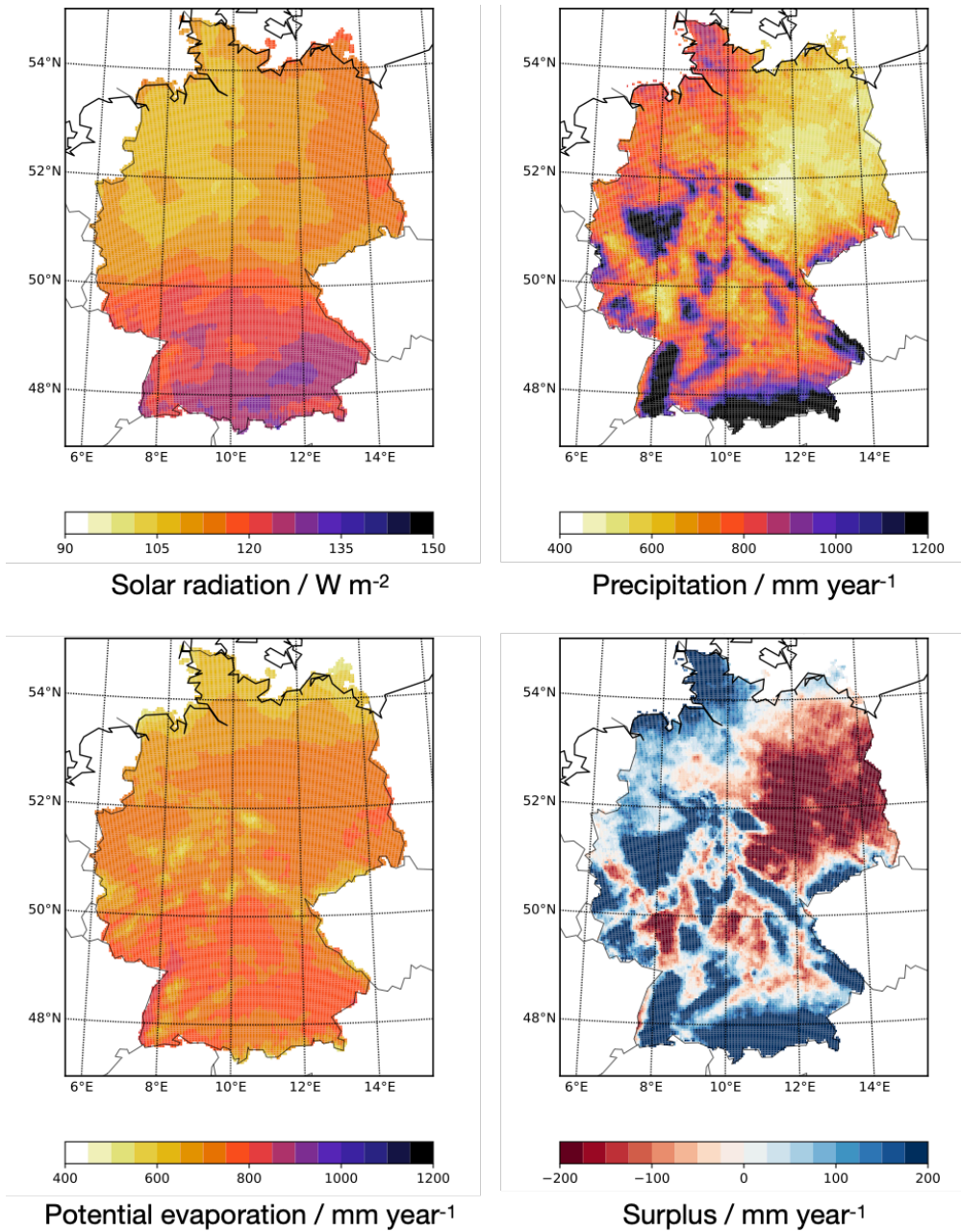

**FIG. 2 WATER AVAILABILITY IN GERMANY**
*Annual average over Germany for the climatological reference period of 1961-1990 of solar radiation (top left), precipitation (top right), potential evaporation (bottom left) and the water surplus (bottom right), i.e. the difference between precipitation and potential evaporation as a measure of water availability. Solar radiation can be converted into an equivalent water flux with the help of latent heat of evaporation with approx. 1 W m$^{-2}$ = 12.5 mm per year: The annual average of 120 W m$^{-2}$ thus corresponds to approx. 1500 mm per year (data source: DWD-HYRAS [4]).*

Figure 2 shows that the distribution of water availability in Germany is quite uneven. It rains considerably more in the west, in the mountain ranges in central Germany and in the Alpine region, while precipitation is considerably lower in the east and in some areas of southern Germany. The input of solar radiation as an energy source for evaporation, on the other hand, is relatively evenly distributed - as is potential evaporation.

This results in a very different distribution of water availability. We get it from the difference between precipitation and potential evaporation. If precipitation is higher than potential evaporation, then there is a surplus of water and runoff can occur, which is transported to the sea via rivers. If, on the other hand, the potential for evaporation is greater than the precipitation, then a region lacks water and evaporation cannot be maintained throughout the year. This results in a surplus of water in the west of Germany, while the east actually lacks water from an atmospheric perspective, meaning it is drier throughout the year.

## Climate change in Germany

So what happens with global climate change? The starting point is the increased greenhouse effect caused by the increased concentration of greenhouse gases. This initially leads to the atmosphere becoming more absorptive of the longwave radiation emitted by the surface. The atmosphere therefore absorbs more and radiates this energy both into space and back to the surface - the downwelling longwave radiation then enhances the heating of the surface. This changes the surface energy balance and therefore also the temperature and energy availability, which determines evaporation.

Let us first look at the observed changes before we interpret them and bring them together with the thermodynamic picture of the hydrological cycle. Figure 3 summarises the changes observed at weather stations in recent decades (see also map S1 and frequency distributions S2 in the supplementary material). The spatial patterns are shown in Figure 4, and the climatological means for the climatological reference period of 1961-1990 and the differences in the period 1991-2020 are summarised in Table 1.

At 1.1 K, Germany's mean temperature has increased significantly more than the global mean temperature (0.5 K) since the reference period. Precipitation, on the other hand, is highly variable. Its annual fluctuations represent a significant proportion of the mean value of 778 mm per year, but does not show a clear trend. At 713 mm per year, potential evaporation is comparable to precipitation, but has been increasing relatively continuously since 1990. Compared to the reference period, it has increased by an average of 7.6% or 54 mm per year in the years 1991-2020. The trends in temperature and potential evaporation have increased significantly and almost linearly since 1990.

The lack of increase in precipitation, contrary to which is actually expected with the strengthening of the hydrological cycle with global climate change, can be explained by the fact that the moisture in Germany comes mainly from the ocean. However, there has only been a slight increase in temperatures there, so evaporation should only have increased slightly so far. With 0.5 K warming over the ocean and an increase in evaporation of 2-3 % $K^{-1}$ [6], one would expect a corresponding increase in Germany of around 778 mm per year · 2.5 % $K^{-1}$ · 0.5 K = 10 mm per year. This is significantly less than the fluctuations in precipitation from year to year. Such trends are therefore not recognisable in the strong fluctuations.

The changes in potential evaporation, on the other hand, are much more pronounced. They can be easily estimated from the changes in the surface energy balance (Figure 3 below). We calculate these from the observed increase in temperature and solar radiation, and determine the partitioning into emission and turbulent heat fluxes by assuming maximum power (see supplementary material).

The changes show both an increase in the downwelling longwave radiation - as expected from global climate change - but also a significant increase in the absorption of solar radiation. The increases in these two warming terms are similarly strong. In the period 1991-2020, they increased on average by $\Delta R_s$ = 5.1 W m$^{-2}$ and $\Delta R_{l,down}$ = 6.1 W m$^{-2}$. This results in increases in net surface emission and turbulent heat fluxes of around $\Delta J$ = 5.6 W m$^{-2}$ each. The potential evaporation has thus increased by an average of 49 mm per year in the period 1991-2020. This fits very well with the 54 mm per year estimated using the empirical approach (Table 1). The main driver of the increasing drought in Germany is therefore predominantly the strong increase in potential evaporation.

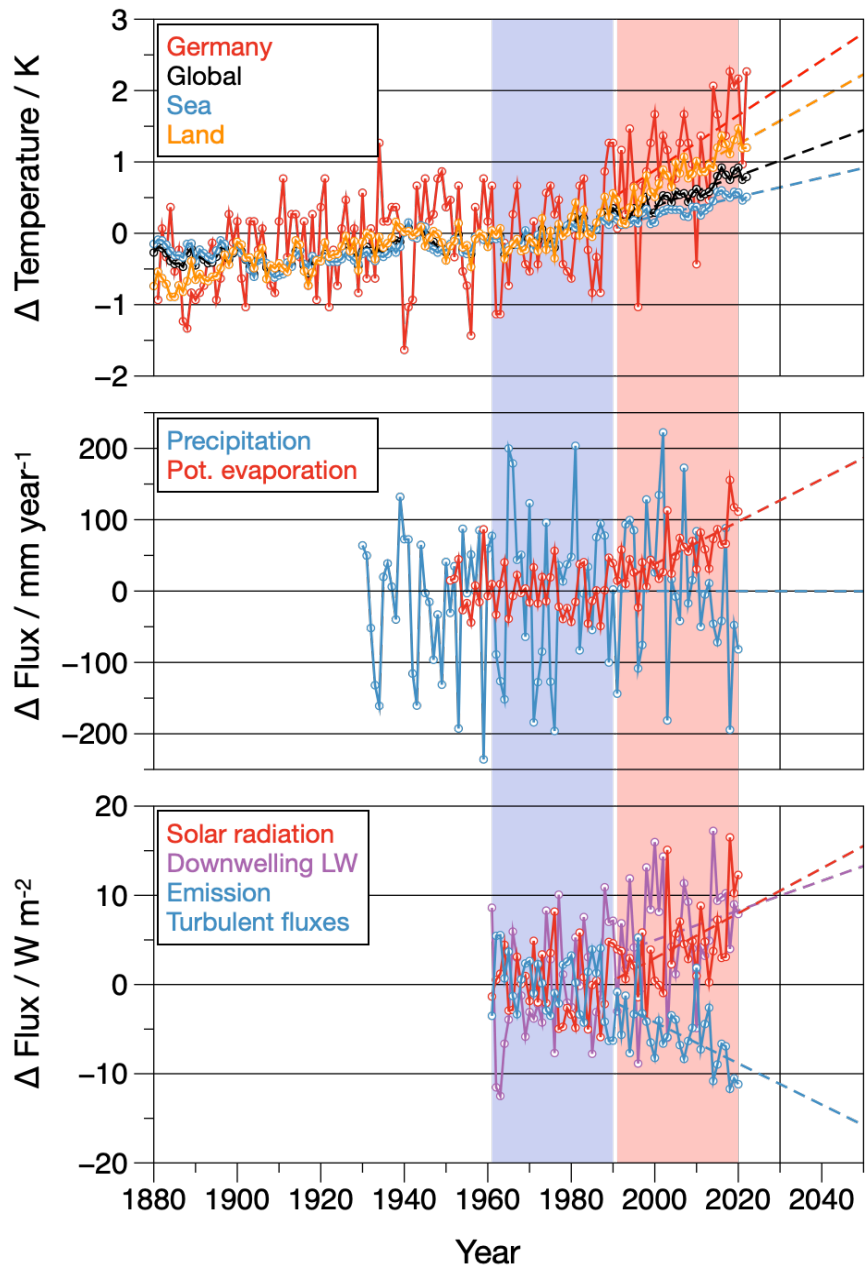

**FIG. 3 CLIMATE CHANGE**
*Time series of observed deviations in annual mean temperature (top) globally (black), over the sea (blue), over land (orange) and over Germany (red), as well as the averaged deviations over Germany from (centre) precipitation (blue) and potential evaporation (red), and (bottom) energy fluxes of the surface energy balance, i.e. warming through absorption of solar radiation (red) and the downwelling longwave (LW) radiation emitted by the atmosphere (purple), as well as cooling through emission and turbulent fluxes (blue). The reference period from 1961-1990 is highlighted in blue, the period 1991-2020 in red (data sources: DWD-HYRAS [4], NASA-GISS).*

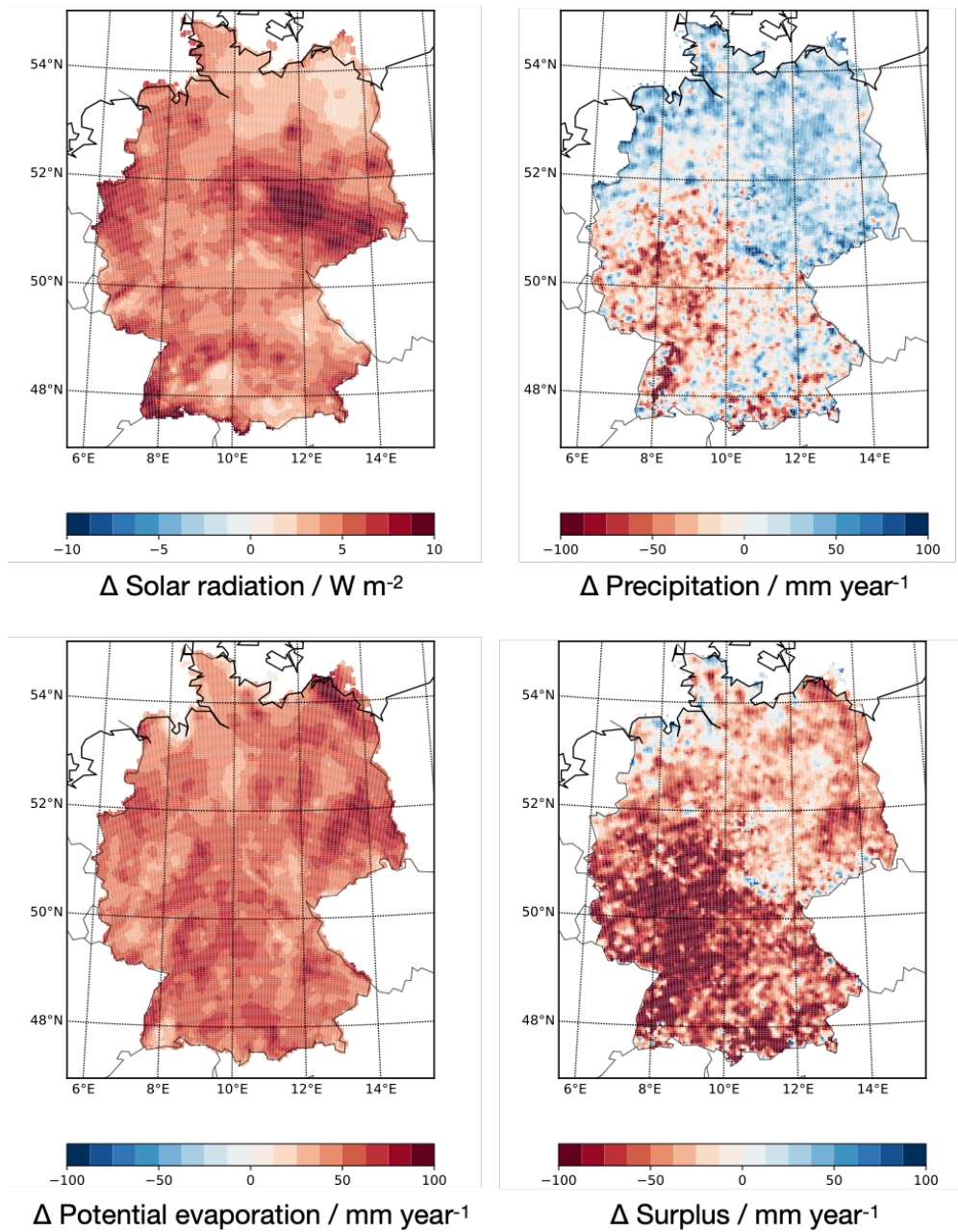

**FIG. 4 CHANGES IN WATER AVAILABILITY**
*Changes in the annual means of solar radiation (top left), precipitation (top right), potential evaporation (bottom left) and water surplus (bottom right) for the period 1991-2020 compared to the climatological reference period (data source: DWD-HYRAS [4]).*

## Acceleration of the hydrological cycle

So how can we explain this increase in solar radiation and the increase in extremes? There can be several reasons for the increase in irradiation - for example, a decrease in aerosols or the shift from precipitation to late summer. However, it can also be explained by the acceleration of the hydrological cycle due to climate change. To do so, we use the thermodynamic model (see

**TAB. 1 MEAN VALUES AND TRENDS IN GERMANY**

|  | Climatological mean 1961-1990 | Mean difference 1991-2020 |
|---|---|---|
| Temperature (°C) | 8,2 | 1,1 |
| Precipitation (mm year$^{-1}$) | 778 | 0 |
| Potential evaporation (mm year$^{-1}$) | 713 | 54 |
| Absorbed solar radiation (W m$^{-2}$) | 115 | 5,1 |

Mean values and trends of hydroclimatological variables for Germany shown in Figures 2 and 3.

"Thermodynamics of the hydrological cycle" and supplemental material) and relate the expected changes in the hydrological cycle to these observations.

The starting point for the expected changes to the hydrological cycle due to climate change is the state of saturation - i.e. thermodynamic equilibrium. At warmer temperatures, the atmosphere can hold more water vapour. This is determined by the temperature dependence of the saturation vapour pressure, which is described by the Clausius-Clapeyron equation. From it we can express this directly as a relative change, it is $1 / e_{sat} \cdot de_{sat} / dT = L / (R_v T^2)$ in % K$^{-1}$ (see supplementary material). With an average temperature of $T$ = 288 K, it yields an increase of 6.5 % per degree K of warming. The saturated atmosphere therefore contains more water vapour and, according to equation 1, the disequilibrium also increases. The phases of humidification and dehumidification of the atmosphere therefore intensify. The water cycle thus becomes more variable.

At 8.7 % K$^{-1}$, evaporation increases at a similar rate to the disequilibrium, so the length of the evaporation period remains roughly the same (see additional material). Precipitation, however, changes significantly more. The additional water vapour in the atmosphere increases the work done, which is driven by the heat of condensation. Not only does the rained-out moisture increase (by 6.5 % K$^{-1}$, according to the Clausius-Clapeyron equation), but the efficiency term also increases, as the surface temperature rises due to the stronger greenhouse effect, whereas the radiative temperature does not. As a result, the work performed increases by 9.2 % K$^{-1}$ more than the saturation vapour pressure.

More work leads to more dehumidified air, which is reflected in the increased disequilibrium, and to more acceleration. The velocity therefore increases by 4.6 % K$^{-1}$. As a result, there are two factors that increase the rate of precipitation: More heat is released during condensation, and moist air is replenished more quickly. This increases the precipitation rate by 11.1 % K$^{-1}$. Precipitation is therefore not only increased in amount, the event is also shorter - the precipitation phase is thus shortened by 4.6 % K$^{-1}$. Dehumidification therefore happens faster, it rains shorter and heavier. This trend is also reflected in observations [7].

This acceleration of the hydrological cycle should have an effect on the solar radiation: if condensation is faster because there is more power, then clouds should live for a shorter time. The solar radiation that reaches the surface should then increase as well. And we have already seen this trend in Figure 3.

On average, clouds reduce solar radiation at the surface considerably. The effect can be seen in the difference between the potential radiation - i.e. the incident solar radiation from space - and the solar radiation at the surface. In Germany, the average potential radiation is around 279 W m$^{-2}$. From this, 20-25 % is lost through absorption by water vapour before it reaches the Earth's surface on a cloud-free day. The remaining 200-220 W m$^{-2}$ are then considerably greater than the mean irradiation at the surface of 138 W m$^{-2}$ (Table 1 shows the mean value of absorbed solar radiation of 117 W m$^{-2}$, which is reduced by 15 % due to the surface albedo). Clouds therefore reduce the average in Germany by 60-80 W m$^{-2}$. If their lifetime decreases by even a few per cent due to the stronger acceleration of the air, this can explain the observed increase in solar radiation at the Earth's surface quite well.

## Conclusions

We have described the hydrological cycle as a system far from thermodynamic equilibrium, where the disequilibrium is expressed in unsaturated air. The cycle can be divided into phases of evaporation, which depletes the disequilibrium, and precipitation, where work is performed that restores the disequilibrium.

Even if this picture is of course very simple and can certainly be improved in many aspects with more details, it is quite easy to understand that the hydrological cycle becomes more powerful, is intensifying and is becoming more variable due to climate change. Air then holds more water vapour, and this is the fuel that generates more motion during condensation. Precipitation events will therefore be stronger and shorter. Evaporation, on the other hand, is limited by the availability of energy at the surface. This shortens periods of precipitation and makes it sunnier. The hydrological cycle is therefore becoming more extreme as a result of climate change. This general insight can be derived directly from the thermodynamic description of the water cycle without the need for much detail.

Meteorological observations in Germany show that it has become significantly drier over the last few decades due to climate change. Precipitation does not show a clear trend, which can be explained by the fact that the oceans have not warmed up much so far. Potential evaporation, on the other hand, has increased significantly - this can be seen both in an empirical estimate and in the changes in the surface energy balance. On the one hand, this is due to the increase in the downwelling longwave radiation - i.e. the stronger greenhouse effect - and an almost equally strong increase in absorbed solar radiation at the surface. This can be explained by the acceleration of the water cycle, because the lifespan of clouds should become shorter due to climate change. As this trend towards greater dryness was generally explained here via thermodynamics, these effects and trends should also be expected in similar climates in neighbouring countries such as the Netherlands, Austria or Poland.

The effects of climate change can be seen quite clearly in Germany, not only in the increase in temperature, but also in increasingly dry conditions and heavier precipitation events.

## Summary

*Droughts and heavy rainfall are caused by strong spatiotemporal differences in evaporation and precipitation. In physical terms, they represent fluxes that deplete and generate a state of thermodynamic equilibrium with regard to the saturation of water vapour in the atmosphere. Climate change intensifies these differences and the disequilibrium because warmer air can hold more moisture. Precipitation events become stronger and shorter, as more condensation generates more power, which leads to more motion and dehumidification. In Germany, these effects are mainly reflected in increased potential evaporation and an increase in the absorption of*

*solar radiation at the surface, while precipitation shows no clear trend. As a result, Germany has already become considerably drier due to climate change.*

# Keywords

Hydrological cycle, drought, evaporation, precipitation, thermodynamics, climate change, dryness, power.

# Additional material

The original supporting material can be found under "Supporting Information" (in German) at https://onlinelibrary.wiley.com/doi/10.1002/piuz.202301697/suppinfo. The translated version is attached at the end of this document.

# The author

Axel Kleidon studied physics and meteorology at the University of Hamburg and Purdue University, Indiana, USA. He completed his doctorate at the Max Planck Institute for Meteorology in 1998 on the influence of deep-rooted vegetation on the climate system. He then conducted research at Stanford University in California and at the University of Maryland. Since 2006, he has headed the independent research group "Theory and Modelling of the Biosphere" at the Max Planck Institute for Biogeochemistry in Jena. His research interests range from the thermodynamics of the Earth system to the natural limits of renewable energy sources.

# Contact

Dr. Axel Kleidon, Max Planck Institute for Biogeochemistry, P.O. Box 10 01 64, 07701 Jena, Germany. akleidon@bgc-jena.mpg.de

## Box: Thermodynamics of the hydrological cycle

A simple, thermodynamic description of the hydrological cycle has three components: the description of the disequilibrium associated with water vapour in air and the division into two phases of how this disequilibrium is generated and depleted (Figure 1): In the evaporation phase, the surface evaporates, water vapour is brought into the atmosphere by buoyancy, the disequilibrium is depleted and the concentration gets closer to saturation. In the precipitation phase, the water vapour condenses and the latent heat released generates work in a "moist" heat engine. This work moves air and dehumidifies the atmosphere. The water vapour rains out and thus creates the disequilibrium in the form of unsaturated air.

### Disequilibrium

The disequilibrium of the hydrological cycle in the atmosphere is shown by how far the water vapour content is from saturation. Near the surface, this is shown by the relative humidity, $r_h$ (in %), which is then below 100 %. To assign this to a quantity of water, we need the specific humidity of air (in kg m$^{-3}$). It is calculated from the density of the air, $\rho$ (in kg m$^{-3}$), the saturation vapour pressure, $e_{sat}$ (in hPa), which is determined by the Clausius-Clapeyron equation and can be determined numerically using the Magnus formula, for example. The saturation vapour pressure is strongly dependent on the temperature. The specific humidity is then given by $R_a / R_v \cdot \rho \cdot r_h \cdot e_{sat} / p_s$, where $R_a$ and $R_v$ are the gas constants of air and water vapour and $p_s$ is the air pressure at the surface. The water vapour in the lower atmosphere is usually well mixed, in which case the specific humidity is roughly constant with altitude. However, the temperature decreases with altitude, typically with the so-called dry adiabatic lapse rate $\Gamma$ (= 9.8 K km$^{-1}$). Thus, air reaches saturation at a certain height $z_a$. The amount of water vapour, $\Delta m_a$ (in kg m$^{-2}$), that is missing for saturation is therefore described by

$$\Delta m_a = (R_a / R_v) \, \rho \, (1 - r_h) \, (e_{sat} / p_s) \, z_a = (T_s / (L \, \Gamma)) \, (1 - r_h)^2 \, e_{sat}(T_s). \qquad (1)$$

where $L$ is the heat of vaporisation of water ($2.5 \cdot 10^6$ J kg$^{-1}$ K$^{-1}$). The ideal gas law ($\rho = p_s / (R_a \, T_s)$) was used in the derivation. The height $z_a$ was determined by equating the temperature difference over this height with the temperature difference to saturation. The disequilibrium and its sensitivity to heating is therefore mainly described by the strong dependence of the saturation vapour pressure on temperature.

### Evaporation phase

Evaporation is mainly described by the energy input by the absorption of solar radiation, combined with the so-called equilibrium partitioning. The starting point for this is the surface energy balance. This balances the heating by absorption of solar radiation and the downwelling longwave radiation with the cooling by surface emission and turbulent fluxes of sensible heat and the energy equivalent of evaporation. The partitioning into cooling by emission and turbulent fluxes is very well described by the maximisation of the power that can be generated by the sensible heat flux and drives vertical air motion (red arrows in Figure 1) [1]. The maximum possible evaporation rate, i.e. the potential evaporation, is also determined by thermodynamics by assuming that the heated air near the surface remains saturated. This leads to the following expression for evaporation, $L \, E$ (in W m$^{-2}$):

$$L \, E = s / (s + \gamma) \, (R_s + R_{l,down} - R_0) / 2, \qquad (2)$$

where $s$ describes the slope of the saturation vapour pressure with temperature (given by the Clausius-Clapeyron equation), $\gamma$ the psychrometric constant (65 Pa K$^{-1}$), $R_s$ the absorbed solar radiation, $R_{l,down}$ the downwelling longwave radiation, and $R_0$ the emission of radiation at a reference temperature at which the emission of the surface is linearised. The factor of 1/2 comes from the maximisation of the power. The fraction $s/(s + \gamma)$ describes the share of evaporation in the turbulent fluxes and increases with temperature. The increase in the greenhouse effect is reflected in the increase in $R_{l,down}$ and in the temperature, which increases the partitioning factor $s/(s + \gamma)$. For these reasons, the potential evaporation increases with warmer temperatures.

We obtain the length of the evaporation period, $\Delta t_e$, by dividing the disequilibrium by the evaporation rate:

$$\Delta t_e = \Delta m_a/E. \qquad (3)$$

### Precipitation phase

Precipitation can occur after air has reached saturation and water vapour condenses. This releases latent heat, which in turn generates power and performs work. The maximum power $G_{max}$ of this condensation-driven heat engine is given by

$$G_{max} = L\,P\,(T_s - T_r)/T_s = \rho\,C_d\,v^3 + R_v\,T_s\,P \log r_h, \qquad (4)$$

where $P$ is the precipitation rate and the efficiency is described by the temperature difference between the surface, $T_s$, and the radiative temperature $T_r$. The radiative temperature is the temperature at which the absorbed solar radiation is emitted into space, $T_r \approx 255$ K.

On the one hand, the power generates the updrafts that brings water vapour into the cloud and determines the rate $P$, which is balanced by friction ($\rho\,C_d\,v^3$ in equation 4, blue arrows in Figure 1). This is described here by a drag coefficient $C_d$ and a wind speed $v$. On the other hand, work is done to dehumidify the air. This is described by the term $R_v\,T_s\,P \log r_h$ in equation 4. This term creates the disequilibrium.

The power dehumidifies the air over a period of time $\Delta t_p$, which is limited by the mass balance, since $E\,\Delta t_e = P\,\Delta t_p = \Delta m_a$. If we assume that the power is maximised that goes into the generation of convective motion, then the resulting optimal disequilibrium $\Delta m_a$ can be determined (see supplementary material),

$$\Delta m_a = L\,(T_s - T_r)/(2\,R_v\,T_s^2 + L\,(T_s - T_r))\,m_{a,max}, \qquad (5)$$

and yields the precipitation rate $P$,

$$P = m_{a,max}\,v\,/\,z_a, \qquad (6)$$

and the duration of the precipitation phase $\Delta t_p$,

$$\Delta t_p = \Delta m_a/P. \qquad (7)$$

Here $m_{a,max}$ is the amount of water vapour in the atmosphere at saturation and $v$ is the velocity obtained from the maximised dissipation.

Equations (1) to (7) represent a complete description of the hydrological cycle. They are consistent with energy and mass balances, thermodynamics and also describe the rates of the dynamics of humidification and dehumidification as a function of temperature. Whilst the drag coefficient is empirical, the relative changes can nevertheless be determined in general terms (see supplemental material).

**DROUGHT | CLIMATE CHANGE**

Why global climate change amplifies hydrological extremes

# Droughts in Germany

## Supporting material for DOI: 10.1002/piuz.202401697

Axel Kleidon

## 1. Creation of the illustrations

Data sets were used to create the maps and time series in the article that are freely accessible.

Figures 2 and 4 were created using the HYRAS data set of the German Weather Service. The version of the data set with daily totals or daily averages was used for this purpose. The data set of precipitation [S1], documented in [S2], global radiation (the incident solar radiation at the Earth's surface) [S3], and the daily minimum temperature [S4], daily mean temperature [S5] and daily maximum temperature [S6], documented in [S7], was used. The data sets were then processed using the CDO commands [S8] and averaged over the time periods.

To create Figure 2, the data sets were averaged over the period 1961-1990. A surface albedo of 0.15 was assumed to calculate the absorbed solar radiation. Daily maximum and minimum temperatures were used to calculate the diurnal temperature range (DTR) needed for the empirical estimation of potential evaporation (see Section 2). This was done on a daily basis and then averaged. The surplus was determined from the difference between mean precipitation and potential evaporation.

To create the time series in Figure 3, the NASA GISTEMP dataset [S9], documented in [S10], was used to show the global temperature changes and averaged over sea and land. The values were adjusted so that the changes for the reference period of 1961-1990 are zero. The time series for the mean values for Germany were calculated from the HYRAS dataset as well as for the values for potential evaporation determined from it. The determination of the changes in energy fluxes at the surface is described in Section 4.

Figure 4 was created by averaging the mean values for the HYRAS data sets over the period 1991-2020. The difference to the mean values from the reference period is then shown.

## 2. Determination of potential evaporation according to Hargreaves

Potential evaporation is estimated by the Hargreaves method [5] from the potential radiation, the mean daily temperature and the DTR. The potential radiation, $R_{pot}$, is the incident radiation at the top of the atmosphere - it depends only on latitude, day and time of the year (averaged over Germany over the year, it is around 279 W m$^{-2}$). It can be determined using standard formulas for calculating the solar zenith angle.

The potential evaporation, $E_{pot}$, is then calculated using (equation 8 in [5])

$$LE_{pot} = 0.0023 \times R_{pot} \times (T_s + 17.8) \times (T_{max} - T_{min}), \tag{S1}$$

where $LE_{pot}$ is given in units of W m$^{-2}$, $R_{pot}$ in W m$^{-2}$, $T_s$ is the near-surface air temperature in °C and $T_{max} - T_{min}$ is the DTR in K. Divided by 29 W m$^{-2}$/(mm day$^{-1}$), the equation gives the potential evaporation in mm day$^{-1}$.

Equation S1 was calculated for each day over the entire length of the HYRAS dataset where daily minimum and maximum temperatures were available. This was then either averaged over the reference period (for Figure 2 and Table 1), averaged over Germany (for Figure 3), and over the period 1991-2020 to determine the difference (for Figure 4 and Table 1).

## 3. Maximum power and precipitation

The heating of the air during condensation performs work. The partitioning of this work into dehumidification and acceleration is determined by maximizing the work that goes into accelerating the air.

The work done by condensation is described by (equations 4 and 6 from the box in the main text)

$$G_{max}\, \Delta t_p = LP \times (T_s - T_r)/T_s \times \Delta t_p = L\, \Delta m_a \times (T_s - T_r)/T_s. \tag{S2}$$

This work is balanced with the frictional losses associated with motion, $D_{fric}$, and the dehumidification work, $G_{dehum}$,

$$G_{max}\, \Delta t_p = (D_{fric} + G_{dehum})\, \Delta t_p, \tag{S3}$$

with

$$D_{fric} = \rho\, C_d\, v^3 \tag{S4}$$

and

$$G_{dehum} = -R_v\, T_s\, P \times \log r_h. \tag{S5}$$

This work balance can first be transformed to obtain an expression for the velocity $v$ (with $P$ determined by equation 6 in the main text)

$$v = m_{a,max}/(\rho\, C_d\, z_a) \times (L\, (T_s - T_r)/T_s + R_v\, T_s \log r_h), \tag{S6}$$

where $r_h = 1 - \Delta m_a/m_{a,max}$. The velocity can then be expressed by

$$v = m_{a,max}/(\rho\, C_d\, z_a) \times (L\, (T_s - T_r)/T_s + R_v\, T_s \log(1 - \Delta m_a/m_{a,max})). \tag{S7}$$

With more dehumidification work, the disequilibrium $\Delta m_a$ increases, the relative humidity $r_h$ decreases and the velocity decreases accordingly. However, the lower velocity then reduces the precipitation rate $P$ and thus the power that goes into motion. Therefore, there is a maximum of work that goes into generating motion.

The work is then maximized via $d(D_{fric} \Delta t_p)/d\Delta m_a = 0$. Using an approximation for the logarithm $((1-x) \times \log(1-x) \approx -x)$, this leads to an optimum value for dehumidification, $\Delta m_{a,opt}$:

$$\Delta m_{a,opt}/m_{a,max} = L\ (T_s - T_r)/(2\ R_v\ T + L\ (T_s - T_r)). \tag{S8}$$

This then sets the value for the relative humidity, $r_h = 1 - \Delta m_{a,opt}/m_{a,max}$, after dehumidification during the precipitation phase.

This results in the following expression for the optimum speed

$$v_{opt}^2 \approx 1/(\rho\ C_d\ z_a) \times L\ m_{a,max}\ (T_s - T_r)/T_s \times (R_v\ T_s + L\ (T_s - T_r)/T_s)/(2\ R_v\ T_s + L\ (T_s - T_r)/T_s)$$

$$\tag{S9}$$

This rate can then be used to determine the precipitation rate $P$ (equation 6) and the duration of the precipitation period $\Delta t_p$ (equation 7) as well as the relative changes with global warming (section 5).

## 4. Energy balance changes in climate change

The changes in the surface energy balance shown in Figure 3 were estimated as follows. The surface energy balance over land on a daily average is described by

$$R_s + R_{l,down} = \sigma\ T_s^4 + H + LE, \tag{S10}$$

where the ground heat flux was neglected here, as it is generally small. Here $R_s$ is the absorption of solar radiation at the surface, $R_{l,down}$ is the downwelling longwave radiation (the greenhouse effect), $\sigma\ T_s^4$ is the thermal radiation of the surface ($\sigma$ is the Stefan-Boltzmann constant, $\sigma = 5.67 \times 10^{-8}$ W m$^{-2}$ K$^{-4}$), and $H$ and $LE$ are the sensible and latent heat flux.

In order to estimate the changes due to climate change, the deviations from the climatological mean are considered and the Stefan-Boltzmann law is linearized. The energy balance changes then follow

$$\Delta R_s + \Delta R_{l,down} = k_r\ \Delta T_s + \Delta H + \Delta LE, \tag{S11}$$

where $k_r = 4\ \sigma\ T_s^3$ is derived from the linearization of the Stefan-Boltzmann law and the climatological mean temperature of $T_0 = 8.2$ °C (Table 1). From the observations we obtain $\Delta R_s$ and $\Delta T_s$. We then assume maximum power and obtain that $\Delta H + \Delta LE \approx (\Delta R_s + \Delta R_{l,down})/2$. We can then solve for the unknown change $\Delta R_{l,down}$ and get:

$$\Delta R_{l,down} = 2\ k_r\ \Delta T_s - \Delta R_s. \tag{S12}$$

With $k_r = 5.1$ W m$^{-2}$ K$^{-1}$, $\Delta T_s = 1.1$ K, and $\Delta R_s = 5.1$ W m$^{-2}$ we thus obtain $\Delta R_{l,down} = 6.1$ W m$^{-2}$. The changes in $\Delta H + \Delta LE$ are therefore $\Delta H + \Delta LE = 5.6$ W m$^{-2}$. At summertime temperatures around $T_s = 20$ °C, the equilibrium partitioning factor $s/(s + \gamma) \approx 0.7$, so that the potential evaporation has increased by around 3.9 W m$^{-2}$.

Converted with 29 W m$^{-2}$/ (mm day$^{-1}$) and integrated over the year, this results in an increase of 49 mm year$^{-1}$.

## 5. Relative changes in the hydrological cycle

We express the changes in the hydrological cycle due to climate change as relative changes per degree of warming, i.e. in units of % K$^{-1}$. This has the advantage that these changes can be calculated in general, the expressions are often shorter, can be easily combined with each other and we do not need some undetermined parameters.

The starting point is the increase in water vapor in the atmosphere at saturation, i.e. the state of thermodynamic equilibrium. It is described by the Clausius-Clapeyron equation. We can transform this equation to calculate the relative change directly:

$$1/e_{sat} \times de_{sat}/dT = L/(R_v T_s). \tag{S13}$$

With a value of the enthalpy of evaporation of $L = 2.5 \times 10^6$ J kg$^{-1}$ K$^{-1}$, the molar gas constant for water vapor of $R_v = 461$ J kg$^{-1}$, and a global mean temperature of $T_s = 15°C = 288$ K, this results in a relative change of 6.5 % K$^{-1}$.

The disequilibrium (equation 1 in the box) is mainly determined by the temperature dependence on the saturation vapor pressure. Even if the temperature appears explicitly in equation 1, the relative temperature change is small (1 K / 288K ≈ 0.3% K$^{-1}$), as absolute temperatures are considered here. The possible change in relative humidity is neglected here (see justification below), so that

$$1/\Delta m_a \times d\Delta m_a/dT_s \approx 1/e_{sat} \times de_{sat}/dT_s \tag{S14}$$

and the disequilibrium therefore also scales with around 6.5 % K$^{-1}$.

The relative change in evaporation with warming due to the greenhouse effect can be described as

$$1/LE \times d(LE)/dT_s = 1/(s/(s + \gamma)) \times d(s/(s + \gamma))/dT_s + 1/(R_s + R_{l,down} - R_0) \times dR_{l,down}/dT_s \tag{S15}$$

and thus consists of two contributions (see also [5]), the relative change in the factor associated with the equilibrium partitioning and the relative change in the warming of the Earth's surface due to the absorption of radiation. The contribution of the first part can be determined as follows: The Magnus formula $e_{sat} = 603$ Pa $\times \exp(17.15\ T/(234.7 + T))$, where the temperature $T$ enters in °C, yields a value of $e_{sat} = 2338$ Pa for $T \approx 20$ °C. This means that $s = de_{sat}/dT = L\ e_{sat}/(R_v T_s) = 147.7$ Pa K$^{-1}$. It results in a distribution factor of $s/(s + \gamma) = 147.7/(147.7 + 65) = 0.694$. The relative change is then around 1.2 % K$^{-1}$. The second term represents the relative change in heating due to absorption. With $R_s = 117$ W m$^{-2}$, $R_{l,down} \approx 320$ W m$^{-2}$, $R_0 = 356$ W m$^{-2}$ (Stefan-Boltzmann emission at the mean annual temperature) and $\Delta R_{l,down} = 6.1$ W m$^{-2}$ we obtain a relative change of 7.5 % K$^{-1}$. Together, this results in a change of 8.7 % K$^{-1}$. This is slightly more than the 6.9 % K$^{-1}$ calculated from Table 1, but is of a similar order of magnitude.

The relative change in the length of the evaporation period is determined by applying the chain rule

$$1/\Delta t_e \times d\Delta t_e/dT_s = 1/\Delta m_a \times d\Delta m_a/dT_s - 1/E \times dE/dT_s. \tag{S16}$$

We can calculate this directly using the relative changes determined above and obtain a decrease of -2.2% K$^{-1}$.

The relative change in power integrated over the precipitation period is calculated from equation 4 as

$$1/(G_{max} \Delta t_p) \times d(G_{max} \Delta t_p)/dT_s = 1/\Delta m_a \times d\Delta m_a/dT_s + 1/(T_s - T_r) \times T_r/T_s. \quad (S17)$$

The first term changes by 6.5% K$^{-1}$ (see above), the second by 2.7% K$^{-1}$ at temperatures of $T_s$ = 288 K and $T_r$ = 255 K. Together, this results in an increase in power of 9.2% K$^{-1}$, the power thus increases more with temperature than the state of saturation.

The relative dehumidification, $\Delta m_{a,opt}/m_{a,max}$, that follows from the maximization and thus the relative humidity, $r_h$, are relatively insensitive to temperature changes. This was determined numerically and results in a sensitivity of around -1% K$^{-1}$, i.e. a slight decrease and slightly more relative dehumidification at higher temperatures.

The relative change in velocity with temperature changes in proportion to the work performed - this can be obtained directly from equation (S8), whereby the last term on the right-hand side changes little with temperature:

$$1/v \times dv/dT_s \approx 1/2 \times 1/(G_{max} \Delta t_p) \times d(G_{max} \Delta t_p)/dT_s. \quad (S18)$$

If the power increases by 9.2% K$^{-1}$, then the speed increases accordingly by around 4.6% K$^{-1}$.

For the relative increase in the precipitation rate (equation 6), we then obtain

$$1/P \times dP/dT_s = 1/m_{a,max} \times dm_{a,max}/dT_s + 1/v \times dv/dT_s. \quad (S19)$$

The relative change in saturation, $1/m_{a,max} \times dm_{a,max}/dT_s$, scales like Clausius-Clapeyron, i.e. with 6.5% K$^{-1}$, the velocity with 4.6% K$^{-1}$. The precipitation rate therefore scales with 11.1% K$^{-1}$. This is significantly stronger than Clausius-Clapeyron, as there is the additional contribution of the increased velocity.

The length of the precipitation phase changes accordingly with warming. With equation (7) this results in

$$1/\Delta t_p \times d\Delta t_p/dT_s = 1/\Delta m_a \times d\Delta m_a/dT_s - 1/P \times dP/dT_s. \quad (S20)$$

With the relative changes in disequilibrium, which scales with Clausius-Clapeyron, and precipitation, which scales more strongly than Clausius-Clapeyron due to the increased velocities, we thus obtain a shortening of 4.6 % K$^{-1}$ (i.e. a decrease, -4.6 % K$^{-1}$). This decrease is caused by the acceleration work, as it increases more than the saturation content of the atmosphere.

## 6. Temperature differences between the years 1991-2020 and the climatological reference period

As additional information, Figure S1 shows the increase in mean annual temperature between 1991-2020 and the climatological reference period of 1961-1990 of the HYRAS data set.

Figure S2 shows the temperature increase at four selected DWD weather stations: that of the author's place of residence (Jena), a non-urban weather station (Fichtelberg - the highest mountain in the Erzgebirge), a location in the north (Hamburg) and in the south (Munich). The shift of the frequency distributions of around 1.2 K between the period 1991-2020 and the climatological reference period show the same order of magnitude as the increase shown in Figure S1. Also shown are the frequency distributions of the last five years, which show a significantly stronger warming and match the trends in Figure 3 in the main text.

**FIG. S1 TEMPERATURE INCREASE IN GERMANY**

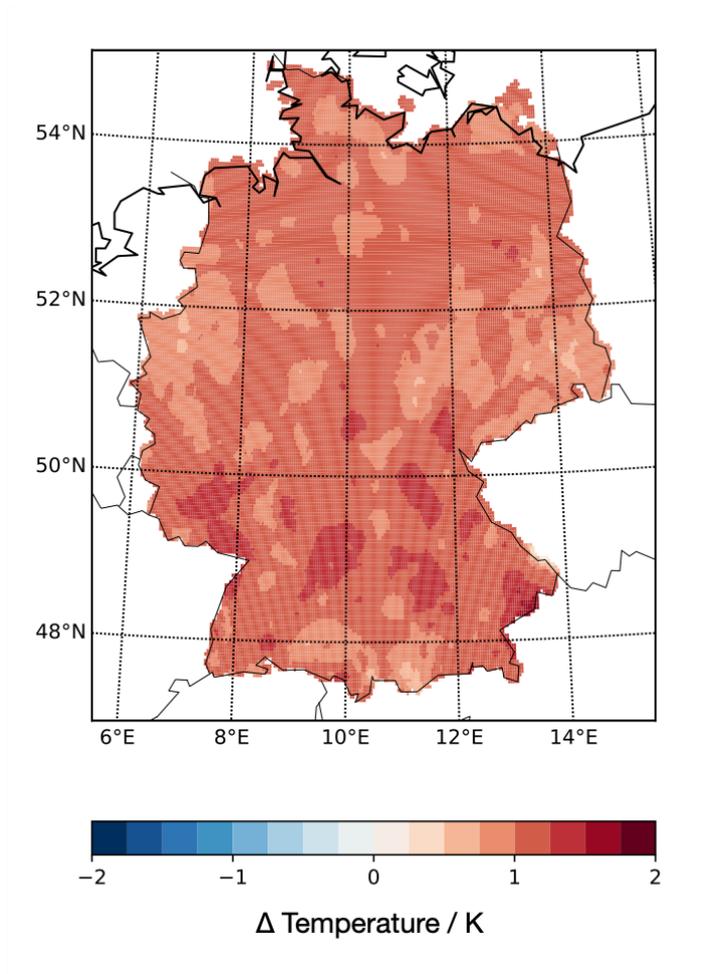

*Difference in the annual mean daily temperature for Germany for the years 1991-2020 compared to the climatological reference period of 1961-1990 (data source: DWD-HYRAS [4]).*

**FIG. S2 TEMPERATURE INCREASE AT WEATHER STATIONS**

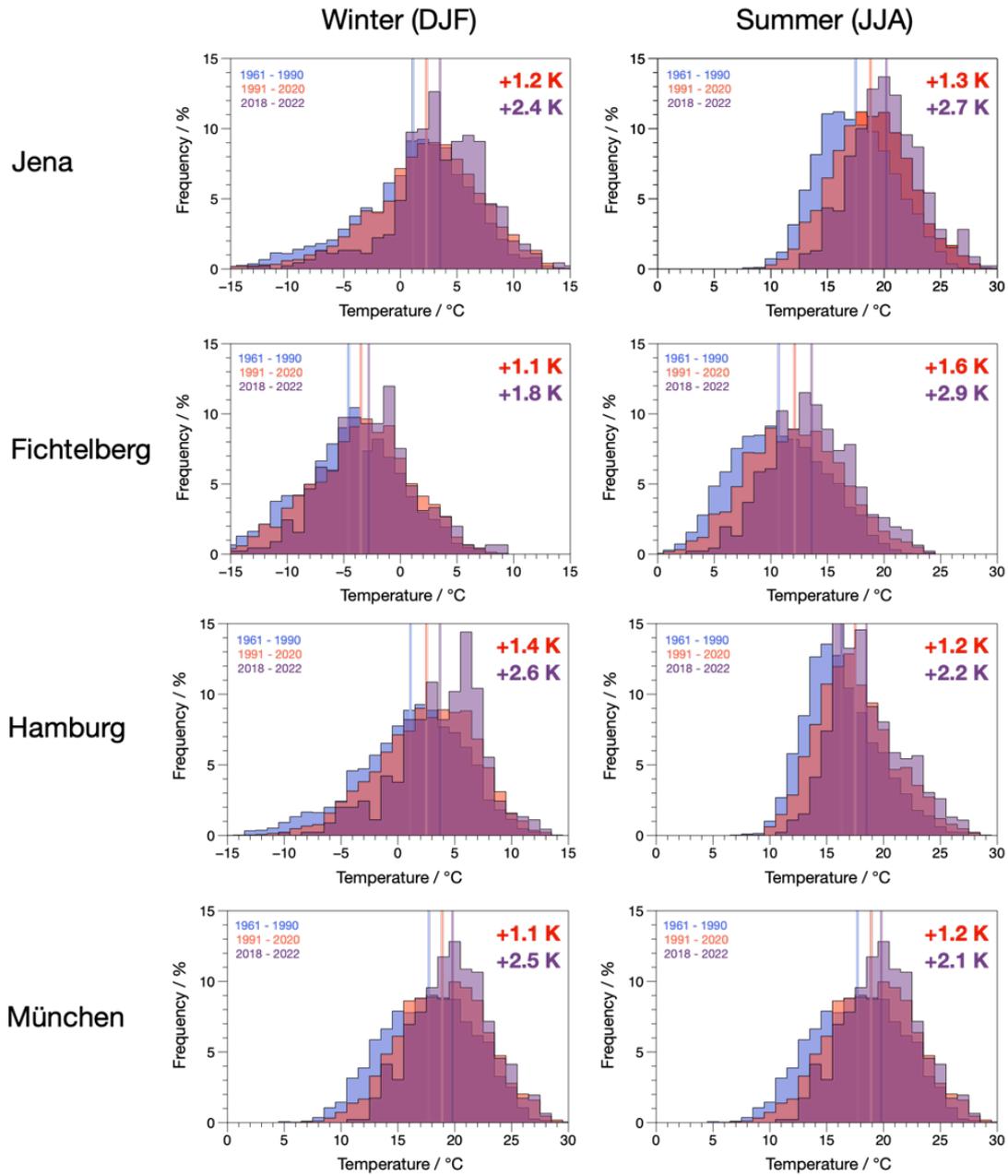

*Frequency of daily mean temperatures at four German weather stations - Jena Observatory (station ID 02444), Fichtelberg (ID 01358), Hamburg-Fuhlsbüttel (ID 01975), Munich-Bavariaring (ID 03379) - for winter (left, December-February) and summer (right, June-August). The observations for the climatological reference period of 1961-1990 are shown in blue, for the period 1991-2020 in red, and those for the last five years in purple. The figures shown are the mean changes in the frequency distributions (data source: DWD).*